# Knowledge-Based Legal Document Assembly

**Marko Marković[1], Stevan Gostojić[1]**

[1]University of Novi Sad, Faculty of Technical Sciences, Serbia

{markic,gostojic}@uns.ac.rs

## Abstract

This paper proposes a knowledge-based legal document assembly method that uses a machine-readable representation of knowledge of legal professionals. This knowledgebase has two components – the formal knowledge of legal norms represented as a rule-base and the tacit knowledge represented by a document template. A document assembly system is developed as a proof of concept. It collects input data in the form of an interactive interview, performs legal reasoning over input data, and generates the output document. The system also creates an argument graph as an explanation of the reasoning process providing the user with an interpretation of how the input data and the rule-base influence the content of the output document. The system also semantically marks up data in the output document, facilitating further processing and providing support to the interoperability of information systems in the legal domain.

## 1 Introduction

People participating in judicial and administrative proceedings are expected to have a certain knowledge to communicate with government authorities efficiently. They usually have to hire a legal counselor to represent them in these proceedings. Law students and young legal professionals may also face challenges when assembling documents in which content is regulated by law.

Legislation prescribes the content and the structure of some legal document types but gives some freedom to document authors to adapt them to a particular case. For non-lawyers, this vagueness becomes an obstacle as they lack the knowledge needed to compose the document with no strict directions. Often, a solution in this situation is hiring a lawyer to assemble the legal document.

After formal education, legal professionals at the beginning of careers usually perform document assembly tasks with the help of their older colleagues. Also, documents assembled by experienced lawyers serve as document models enabling younger lawyers to create documents by making necessary modifications to existing documents. This method can simplify assembly tasks regardless of how much experience some lawyer has.

Anderson and Manns (2017) argue that lawyers usually create new documents by editing some of the previously created documents and call this common practice "precedent-based drafting". This way, lawyers reuse knowledge and experience embedded in "preceding" documents to save time and money.

The relevant legal document assembly experts' knowledge could be divided into formal knowledge received through education and tacit knowledge gained through experience.

Usually, legal document assembly is supported by text processing software. The computer support could be improved if the legal experts' knowledge was available in a machine-readable format by developing a document assembly system. However, legal knowledge formulated by legislation is in a human-readable format. Therefore, the transformation of legislation into a machine-readable legal rule format is necessary. This format has to support non-monotonic features of legal rules. A document template in a machine-readable format is also needed to capture the tacit knowledge of composing legal documents.

For non-lawyers, the document assembly process should not behave as a black box. Instead, it should provide some explanation for its result enabling users to understand the connection between input data and the output document. The explanation of the assembly process could be helpful for young lawyers as well.

The rest of this paper is organized as follows. The next section gives an overview of related research. The Method section explains the legal document assembly method. The Results section presents the prototype of a legal document assembly system. The last section sums up the benefits of the proposed method and gives some directions for future research.

## 2 Related Work

This section reviews research focused on legal document assembly and legal knowledge representation. The first part discusses the features of several integrated document assembly systems. The second part analyzes available formats suitable for representing legal document assembly knowledge.

There are numerous document assembly systems and some of them, applicable to legal documents, are discussed below.

HotDocs (2020) is software for generating legal documents that usees document templates. It includes a HotDocs Developer, a special tool for creating document templates. Document templates combine textual fragments, variables, and business logic. The software enables the creation of interviews for gathering input data. HotDocs integrates into a word processor and uses answers gathered in the interview and a document template to generate a document. However, logic expressions in HotDocs document templates do not support non-monotonic features typical for legal reasoning.

ContractExpress (2020), formerly DealBuilder, also creates legal documents using document templates. The software is also embedded in a word processor to enable the creation of document templates. Created document templates are used to dynamically build web-based questionnaires for collecting input data. ContractExpress generates a document using a questionnaire and a document template. ContractExpress support rules in document templates using an expression language. This language does not support the non-monotonicity of legal rules.

Pathagoras (2020) is software for assembling documents using so-called 'plain text' document templates. Pathagoras embeds in a word processor and uses specialized markup language for variables and expressions in document templates. The software provides a set of clauses organized by topic, enabling a user to build a document choosing the clauses. Interviews for gathering input data are automatically created. Besides its ease of use, Pathagoras does not support the complexity of rule bases especially non-monotonic set of rules.

XpressDox (2020) is a document automation software based on document templates. It supports integration with databases, web services, and other APIs. XpressDox is available as an add-on for a word processor. It automatically generates an interview based on document templates. Document templates support conditional logic and numerous commands and functions. Logic expressions in XpressDox document templates do not support non-monotonic features of legal rules.

Access to Justice Author (2020) is a software tool that helps non-lawyers to self-represent in court. It enables non-technical persons to design document templates and interactive interviews. The software contains a component called A2J Guided Interview for gathering data in a user-friendly environment. Access to Justice Author supports using variables and macros in document templates. Still, the software in its expressions does not support non-monotonicity necessary to represent rules in legal norms.

DocuPlanner (Branting *et al.*, 1997) is a software system for generating legal documents. It is based on the fact that legal reasoning and legal drafting are interconnected tasks and applies a concept called "queryable liveness" enabling the user to find out how some text region originated from input data. This gives the user an insight into the reasoning behind the assembly process. DocuPlanner does not support the design of interviews for gathering facts.

Rule engines with potential for application in the legal domain are described below.

SPINdle (2020) is a logical reasoner with support for defeasible reasoning. It is developed in Java and exists in two forms, as a standalone reasoning tool, and as a Java library. It supports two file formats for the rule base. One format is based on XML, and another called DFL uses plain text syntax. SPINdle supports modal defeasible logic which makes it applicable in the legal domain. Still, SPINdle doesn't support theory grounding.

DR-DEVICE (Bassiliades *et al.*, 2004) is a defeasible reasoning reasoner for the Semantic Web. It is based on the CLIPS production rules and support rules in an extension of RuleML format called DR-RuleML (Kontopoulos *et al.*, 2011). During reasoning, DR-DEVICE transforms the rule base into CLIPS language and use it with facts represented by an RDF document. After reasoning, conclusions are exported in RDF format and their proofs in an XML document (Bassiliades *et al.*, 2007).

Prova is a rule engine developed in Java (Kozlenkov and Paschke, 2020). It combines Prolog syntax with object-oriented programming. Because Prova is based on Prolog it does not support non-monotonic features of rules. Extending features of Prova is enabled using additional libraries. An implementation of defeasible logic is a part of the ContractLog framework that provides a set of libraries for contractual business logic development (Paschke *et al.*, 2005).

Some rule representation standards are considered for the representation of legal norms in a machine-readable format.

RuleML (Athan *et al.*, 2015a) is an XML format for knowledge representation and rule interchange. The family of RuleML languages is developed to support specific requirements in different application areas. However, RuleML does not support non-monotonic reasoning.

LegalRuleML (Athan *et al.*, 2015b) is developed by OASIS LegalRuleML Technical Committee (OASIS, 2020a) as a rule interchange language. It is an extension of RuleML providing features specific to the legal domain (norms, guidelines, policies, and reasoning). LegalRuleML is suitable for representing legal norms, but it lacks an inference engine.

To represent the structure and to enable generating legal documents, document template formats ToXgene (Barbosa *et al.*, 2002) and FreeMarker (2020) are considered.

ToXgene is a template-based XML document generator. It provides a simple Java API for document generation in Java applications (ToXgene, 2020). Syntax of ToXgene templates called Template Specification Language (TSL) is a subset of the XML Schema language. TSL offers control over the structure and content of generated documents. ToXgene is able to generate random elements, attributes, and content based on several probability distribution types. ToXgene library contains sample lists of first names, last names, country names, city names, etc.

FreeMarker is an open-source template engine developed in Java. It uses a syntax called FreeMarker Template Lan-guage (FTL) for template design that makes templates easy to read and understand. There are several development tools providing support for modeling FreeMarker templates.

For document assembly, it is important to choose a legal document format that suits the purpose of document use. Besides general document formats, there are several standards for representation of legal documents e.g. MetaLex (Winkels *et al.*, 2003) and Akoma Ntoso (Palmirani and Vitali, 2011).

CEN MetaLex is an XML schema for the representation of legal documents (Boer and Winkels, 2011). CEN Meta-Lex supports schema extension, adding metadata, naming, cross-referencing, and constructing compound documents. It is a generic standard that aims to be the lowest common denominator of other document standards in the legal domain.

The Akoma Ntoso schema has been proven suitable to describe the structure of judgments in the Serbian judiciary (Marković *et al.*, 2014). Although indictments and motions to indict are not part of the Akoma Ntoso specification, it offers methods for extending its original structure to support other document types. For custom document types, which use the Akoma Ntoso vocabulary, a generic element doc is available. It is also possible to meet specific requirements by redefining Akoma Ntoso schema.

## 3 Method

This section explains the method for legal document assembly using a machine-readable representation of relevant knowledge.

Document assembly is based on two phases. The first phase is the preparation of assembly knowledge and the second phase is the generation of documents using the prepared knowledge. We call them the analysis phase and the synthesis phase.

### 3.1 Analysis phase

This phase aims to create a formal representation of document assembly knowledge.

The knowledge related to legal document assembly is divided into two categories. The first type of knowledge is found in legislation and can be presented as a rule base. The second type of knowledge is tacit knowledge representing skill to compose legal documents.

In the formal representation of rule base, defeasible properties of rules play an important role as defeasible reasoning is one of the main characteristics in the legal domain.

An example of defeasible reasoning in document assembly may occur when someone wants to file a document to start a court procedure and needs to determine which court has authority in a particular case. To resolve court jurisdictions in the Serbian judiciary several laws have to be applied. Considering criminal cases The Criminal Procedure Code (2019) gives jurisdiction to the court at which territory an offense is committed. Territorial jurisdictions of Serbian courts are specified by The Law on the Seats and Territorial Jurisdictions of Courts and Public Prosecutor's Offices (2013). For every municipality in Serbia, there is a basic court that has territorial jurisdiction and for every basic court, there is a higher court that has territorial jurisdiction.

The distinction between the jurisdiction of basic courts and higher courts is determined by The Law on Organization of Courts (2018). This law grants jurisdiction to basic courts in criminal proceedings conducted for an offense with a maximum imprisonment penalty of not more than 10 years. If the proceeding is conducted for a criminal offense with a maximum imprisonment penalty of more than 10 years jurisdiction is granted to the higher court. Penalties for criminal offenses are defined by The Criminal Code (2019). Additionally, if the defendant is a minor The Law on Organization of Courts grants jurisdiction to higher courts as an authority in juvenile criminal proceedings.

This rule might be in conflict with rule that grants jurisdiction by the level of imprisonment penalty if the defendant is a minor and criminal proceeding is conducted for an offense with maximum imprisonment penalty of ten years or less. To resolve these contradictory conclusions the principle *lex specialis* is applied giving priority to rule related to minor defendants as a more specific rule.

To make legal reasoning behind the determination of court jurisdiction computable, these rules have to be represented formally. The rules are formally represented using defeasible logic in notation proposed by Nute (2001).

The rule that grants jurisdiction in criminal proceedings to basic courts can be represented as a defeasible rule:

$$r_1: \textit{max\_imprisonment(Offence, X), X} <= 10$$
$$\Rightarrow \textit{jurisdiction\_level(Offence, basic)} \qquad (1)$$

Similarly, the rule that grants jurisdiction in criminal proceedings to higher courts can be represented as:

$$r_2: \textit{max\_imprisonment(Offence, X), X} > 10$$
$$\Rightarrow \textit{jurisdiction\_level(Offence, higher)} \qquad (2)$$

The rule that gives authority to higher courts in juvenile criminal proceedings can be represented as:

$$r_3: \textit{is\_minor(Defendant)}$$
$$\Rightarrow \textit{jurisdiction\_level(Offence, higher)} \qquad (3)$$

To resolve conflicts between conclusions of rules $r_1$ and $r_3$ a superiority relation can be modeled as:

$$r_1 \prec r_3 \qquad (4)$$

These rules and relations between them are depicted in Figure 1 using an approach for the visual representation of defeasible rules proposed in (Kontopoulos *et al.*, 2008). Boxes with atomic formulas are enclosed in literal boxes consisted of two parts – the upper part for positive atomic formula and the lower part for negative atomic formula. Rules are represented by circles with rule identifiers and arrows that connect rule identifiers with their premises and conclusions. The arrow is pointing to the upper part of the box if the head of the rules is positive. Otherwise, it is point-ing to the lower part of the box. Superiority relation is represented by a sequence of greater-than symbols between rule identifiers.

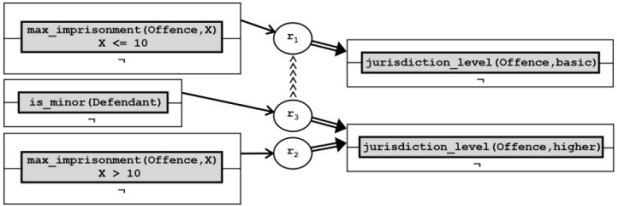

**Figure 1 - Representation of defeasible rules**

To represent rule-base in the machine-readable format we choose LegalRuleML, a format specifically designed for representation of legal norms.

Using the syntax of LegalRuleML rule $r_1$ can be modelled as a <PrescriptiveStatement> element in Listing 1.

---

**Listing 1.** Prescriptive statement for rule $r_1$

```
<lrml:PrescriptiveStatement key="ps_loc_art22para1">
  <ruleml:Rule key=":loc_art22para1"
               closure="universal"
               strength="defeasible">
    <ruleml:if>
      <ruleml:And>
        <ruleml:Atom>
          <ruleml:Rel iri=":max_imprisonment"/>
          <ruleml:Var type=":offence">Offence</ruleml:Var>
          <ruleml:Var type=":years">X</ruleml:Var>
        </ruleml:Atom>
        <ruleml:Atom>
          <ruleml:Expr>
            <ruleml:Fun><=</ruleml:Fun>
            <ruleml:Var>X</ruleml:Var>
            <ruleml:Ind>10</ruleml:Ind>
          </ruleml:Expr>
        </ruleml:Atom>
      </ruleml:And>
    </ruleml:if>
    <ruleml:then>
      <ruleml:Atom>
        <ruleml:Rel>jurisdiction_level</ruleml:Rel>
        <ruleml:Var type=":offence">Offence</ruleml:Var>
        <ruleml:Ind type=":level">basic</ruleml:Var>
      </ruleml:Atom>
    </ruleml:then>
  </ruleml:Rule>
</lrml:PrescriptiveStatement>
```

---

Similarly, the rule $r_2$ can be represented as another <PrescriptiveStatement> element (Listing 2).

---

**Listing 2.** Prescriptive statement for rule $r_2$

```
<lrml:PrescriptiveStatement
                key="ps_loc_art23para1point1">
  <ruleml:Rule key=":loc_art23para1point1"
```

---

```
               closure="universal"
               strength="defeasible">
    <ruleml:if>
      <ruleml:And>
        <ruleml:Atom>
          <ruleml:Rel iri=":max_imprisonment"/>
          <ruleml:Var type=":offence">Offence</ruleml:Var>
          <ruleml:Var type=":years">X</ruleml:Var>
        </ruleml:Atom>
        <ruleml:Atom>
          <ruleml:Expr>
            <ruleml:Fun>></ruleml:Fun>
            <ruleml:Var>X</ruleml:Var>
            <ruleml:Ind>10</ruleml:Ind>
          </ruleml:Expr>
        </ruleml:Atom>
      </ruleml:And>
    </ruleml:if>
    <ruleml:then>
      <ruleml:Atom>
        <ruleml:Rel>jurisdiction_level</ruleml:Rel>
        <ruleml:Var type=":offence">Offence</ruleml:Var>
        <ruleml:Ind type=":level">higher</ruleml:Var>
      </ruleml:Atom>
    </ruleml:then>
  </ruleml:Rule>
</lrml:PrescriptiveStatement>
```

The rule $r_3$ related to juvenile criminal proceedings can be represented as in Listing 3.

---

**Listing 3.** Prescriptive statement for rule $r_3$

```
<lrml:PrescriptiveStatement
                key="ps_loc_art23para1point3">
  <ruleml:Rule key=":loc_art23para1point3"
               closure="universal"
               strength="defeasible">
    <ruleml:if>
      <ruleml:Atom>
        <ruleml:Rel iri=":defendant_is_minor"/>
        <ruleml:Var type="dd:defendant">Defendant</ruleml:Var>
      </ruleml:Atom>
    </ruleml:if>
    <ruleml:then>
      <ruleml:Atom>
        <ruleml:Rel>jurisdiction_level</ruleml:Rel>
        <ruleml:Var type=":offence">Offence</ruleml:Var>
        <ruleml:Ind type=":level">higher</ruleml:Var>
      </ruleml:Atom>
    </ruleml:then>
  </ruleml:Rule>
</lrml:PrescriptiveStatement>
```

---

LegalRuleML provides element <OverrideStatement> to define superiority relations between prescriptive statements. Giving priority to prescriptive statement *ps_loc_art23para1point3* over prescriptive statement *ps_loc_art22para1* is shown in Listing 4.

**Listing 4.** Superiority relation in LegalRuleML

```
<lrml:OverrideStatement>
  <lrml:Override
          under="#ps_loc_art22para1"
          over="#ps_loc_art23para1point3"/>
</lrml:OverrideStatement>
```

The tacit knowledge in document assembly may be described as a skill of composing the document using document fragments. To formally represent these document fragments and how they form the document layout, document templates are used. We have chosen the ToXgene format for document templates because it is based on a well-known XML Schema standard.

To generate a document, ToXgene needs input data. ToXgene supports loading data stored in XML format as name-value pairs. Once loaded, the input data is available in the document template. To place an input value inside a text element ToXgene provides element <tox-sample> for retrieval of data by its name and element <tox-expr> to embed the data value in document. An example of a text fragment in ToXgene format that creates a paragraph containing the name and the birth date of a person is shown in Listing 5.

**Listing 5.** A document fragment in ToXgene format

```
<element name="p">
  <complexType>
    <tox-value>against </tox-value>
    <tox-sample path="[fact_list/fact]"
            where="EQ([name],'defendant')"
            duplicates="no">
      <tox-expr value="[value]" />
    </tox-sample>
    <tox-value>, from </tox-value>
    <tox-sample path="[fact_list/fact]"
            where="EQ([name],'defendant_birthdate')"
            duplicates="no">
      <tox-expr value="[value]" />
    </tox-sample>
    ...
  </complexType>
</element>
```

Besides composing the document using only data entered by the user, reasoning results should also be available in the document template. Therefore, conclusions generated by a reasoner are forwarded to the document template as name-value pairs using the same XML file that contains data entered by the user.

The assembly configuration designed in the analysis phase contains a reference to the rule-base, a reference to the document template, and an interview. The interview includes questions the user should be asked to enter data relevant for document assembly. The interview consists of steps where every step contains the name of the name-value pair and the text of the questions related to the legal fact. The configuration is stored in an XML document.

### 3.2 Synthesis Phase

The purpose of the synthesis phase is to generate the document using configuration created in the analysis phase. This phase consists of an interview, legal reasoning, and document assembly steps.

An interview aims to gather facts needed for document assembly. In the interview, the user is asked to answer a series of questions. Answers to these questions are legal facts relevant to the content of the document.

The reasoning is performed using collected facts and the rule-base proven to be true and the rule-base to obtain its legal consequences. Reasoning conclusions are used in the following assembly steps for generating a legal document and an argument graph.

The template engine generates a legal document using a document template and collected and inferred data. The output format of the document is determined by the document template.

An argumentation graph is generated using proofs for reasoning conclusions. The graph nodes are predicates from the rule-base proven to be true and rules implying that conclusion. The graph edges connect predicates to the rules they appear in.

The Assembly process is repeated in cycles until all interview questions are answered. The flow of the synthesis phase is graphically presented in Figure 2.

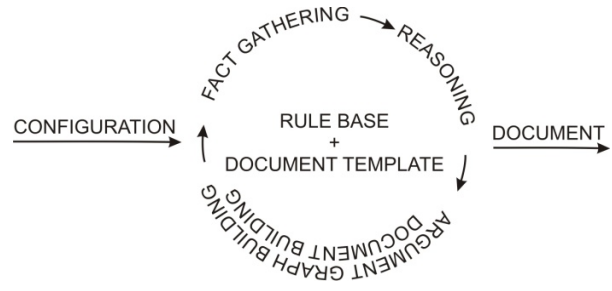

**Figure 2 - Flow of the synthesis phase**

## 4 Results

This section presents a document assembly system developed as a proof of concept for the document assembly method proposed in this paper. The prototype consists of two components – the configuration component and the document assembly component. The configuration component performs tasks that belong to the analysis phase and the document assembly component completes tasks from the synthesis phase.

The configuration component is a software tool that helps a user to design an interactive interview and to connect it with a preexisting rule-base and document template. This tool lets a user choose a predicate from the rule-base that most closely relates to the subject matter of a document. The software then uses this predicate to search the rule base for all other predicates it depends on. It is performed by collecting all predicates acting as premises in rules having the cho-sen predicate as a conclusion. The search continues collecting more predicates that act as premises in rules having any of already collected predicates as a conclusion. The search ends when all predicates that directly or indirectly imply on the chosen predicate are collected.

A user is then asked to assign an interview question to every predicate in the collected set. The order of questions in the interview is determined by the order in which questions are assigned to predicates.

The document assembly component is a web application. It uses configuration created in the previous phase to guide a user through a question answering process. A user is asked a series of questions. After each answer, the system executes the assembly process to show a user content of the document generated by currently collected data. At the same time, the system generates an argument graph as an interpretation of legal reasoning involved in the assembly process.

A user has the ability to return to one of the previous steps at any given time.

The user interface of the document assembly system is shown in Figure 3. The user interface consists of following elements: (A) the progress of question-answering process (a user can choose previous questions to revise previous answers); (B) the pop-up window showing the current question and the appropriate input fields; (C) the document draft based on the facts the user gave so far; (D) the argumentation graph; and (E) the additional explanations related to the current question such as propositions from legislation and excerpts from scholarly texts. As a user progresses through the question answering process, the document content and the argument graph are simultaneously populated on the basis of answered questions. Simple data validation is per-formed after answering each question and feedback is given to a user if the answer is not of the correct data type. The source code of the proposed legal document assembly system is available online at (Legal Document Assembly System, 2020).

## 5 Discussion and Conclusions

The legal document assembly method proposed in this paper uses the knowledge relevant to document assembly in a machine-readable format to generate legal documents. The method is based on two types of knowledge – formal knowledge and tacit knowledge. Formal knowledge is represented by legal norms in a machine-readable format, I.e. as a set of legal rules in LegalRuleML. Tacit knowledge is represented by a document template in ToXgene format.

To demonstrate the applicability of this method for assembling legal documents, the document assembly system is developed. For the provided rule base and document template, the system helps a user to create a document assembly configuration. The system then uses this configuration to ask a user a series of questions in the form of an interview. On every given answer, the system generates the document from input data collected so far.

The system also provides an argument graph as an interpretation of the reasoning process behind document assembly. The argument graph helps the user to understand how rule-base and input data influence the assembly process.

The method is applicable to a variety of legal document types. The assembly of judicial documents that involve the application of several laws and have a complex document structure are good candidates for the application of the pro-posed method. To evaluate the prototype of the assembly system, a group of law students was asked to use the system to generate indictments and give feedback on the utility of the software. A survey completed by 24 students gives an average score of 5 (out of 5) for ease of use and a score of 4.88 (out of 5) for simplification of the assembly process.

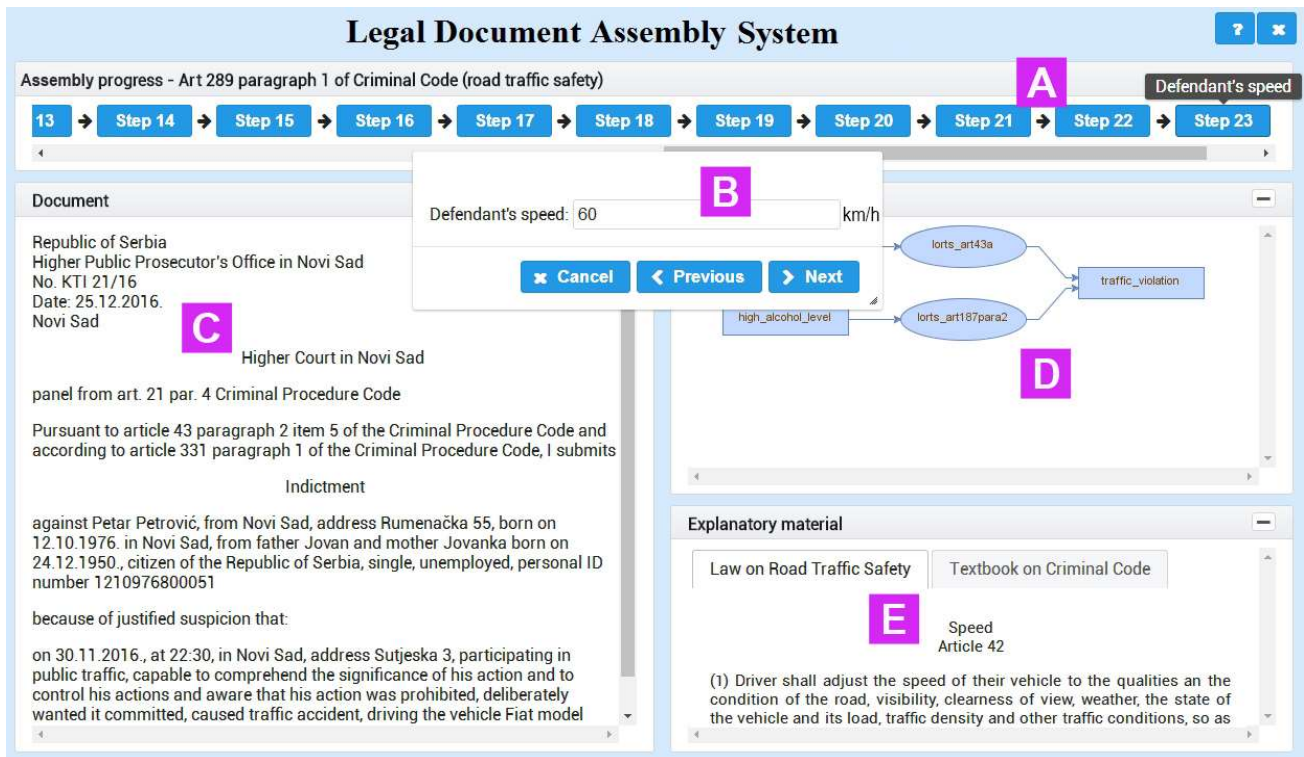

**Figure 3 - User interface of the document assembly system**

The main contribution of this paper is using machine-readable knowledge in legal document assembly. The paper proposes formats for representation of relevant assembly knowledge and a method that applies this knowledge to support document assembly tasks.

In contrast to other solutions in legal document assembly discussed in this paper, the proposed method is based on a machine-readable rule-base format. The rule-base can be developed separately from the document template and one rule-base can be used for assembly of several types of documents. When legislation changes this makes it easier to update the assembly process as a single rule-base has to be kept up-to-date. Besides, this method supports defeasible rules and defeasible reasoning. This is very important for rule representation in the legal domain.

Furthermore, the document assembly system presented in this paper generates documents in Akoma Ntoso standard (Palmirani and Vitali, 2011), a format adopted by OASIS as a basis for the development of LegalDocML standard (OA-SIS, 2020b). Using Akoma Ntoso format the system generates legal documents with structural and semantic markup that improves automatic processing of generated documents. This is beneficial for the interoperability of legal information systems.

An advantage of the proposed document assembly method is the creation of an argument graph. This could be helpful for users with no experience in legal document assembly. Visually represented argument graph shows relations between input data and generated claims in the document. Furthermore, the document assembly system presented in this paper provides the user with explanatory material for a better understanding of the document assembly process. These features give potential to the proposed system for application in the education of future lawyers.

Although the document assembly system provides semi-automatic interview design it does not support the creation of a rule-base and a document template. It should be manually prepared using some XML editing software. This task could be facilitated by the introduction of a domain-specific language (DSL) for formal representation of assembly knowledge. In further research, we plan to tackle this problem.